\documentclass[twocolumn,showpacs,showkeys,preprintnumbers,amssymb,aps,superscriptaddress,pre]{revtex4}
\usepackage{graphicx}
\usepackage{dcolumn}
\usepackage{bm}
\usepackage{epsfig}
\usepackage{color}

\begin{document}

\title{Is order-by-disorder present or absent in a highly frustrated region of the spin-1/2 Ising-Heisenberg model on triangulated Husimi lattices?}
\author{Jozef Stre\v{c}ka}
\email{jozef.strecka@upjs.sk}
\affiliation{Department of Theoretical Physics and Astrophysics, Faculty of Science, P. J. \v{S}af\'{a}rik University, Park Angelinum 9, 040 01, Ko\v{s}ice, Slovak Republic}
\author{Cesur Ekiz}
\email{cekiz@adu.edu.tr}
\affiliation{Department of Physics, Faculty of Science and Letter, Adnan Menderes University, 09010 Ayd\i n, Turkey}

\date{\today}

\begin{abstract}

The geometrically frustrated spin-1/2 Ising-Heisenberg model on triangulated Husimi lattices is exactly solved by combining the generalized star-triangle transformation with the method of exact recursion relations. The ground-state and finite-temperature phase diagrams are rigorously calculated along with both sublattice magnetizations of the Ising and Heisenberg spins. It is evidenced that the Ising-Heisenberg model on triangulated Husimi lattices with two or three inter-connected triangles-in-triangles units displays in a highly frustrated region a quantum disorder irrespective of temperature, whereas the same model on triangulated Husimi lattices with a greater connectivity of triangles-in-triangles units exhibits at low enough temperatures an outstanding quantum order due to the order-by-disorder mechanism. The quantum reduction of both sublattice magnetizations in the peculiar quantum ordered state gradually diminishes with increasing the coordination number of underlying Husimi lattice.

\end{abstract}


\maketitle

\section{Introduction}

Over the last few decades, two-dimensional (2D) frustrated Heisenberg models have become subject of intense research studies owing to an exceptional diversity in a low-temperature magnetic behavior they display \cite{Lhu01,Mis04,Ric04,Lac11}. However, the rigorous treatment of 2D frustrated Heisenberg models is mostly unattainable due to extraordinary difficulties, which bear a close relation to non-commutative (quantum) character of spin operators. Many semi-classical 2D Ising spin systems with competing interactions may also exhibit intriguing cooperative phenomena originating from a geometric spin frustration, but their experimental realizations are regrettably scarce \cite{Jongh,Wolf} at the expense of exact solvability of the respective models \cite{Liebmann86,diep04}. Generally, the spin frustration is indispensable ground for emergence of several remarkable cooperative phenomena such as the \textit{order-by-disorder}, which refers to a selection of unusual ordered state of some highly frustrated classical or quantum spin model driven either by thermal or quantum fluctuations \cite{vil77,vil80,she82,hen89}. Another striking phenomenon closely related to the geometric spin frustration is a reentrance, which occurs if an increase in temperature causes a phase transition from a disordered state emerging at lower temperatures to a spontaneously ordered state emerging at higher temperatures \cite{diep04}. The reentrant phase transitions have been experimentally detected in various magnetic systems including spin glasses \cite{Binder} or even non-magnetic systems such as liquid mixtures \cite{nar94}. 

Exactly solved models belong to the most attracting issues of statistical mechanics, because they provide results not affected by any approximation \cite{baxt82}. It is well known that the exact solution of the Ising model on the Bethe lattice can alternatively be viewed as the rather reliable (Bethe-Peierls) approximation of this model on a regular planar lattice with the same coordination number \cite{Kurata,Katsura}. However, the Ising model on the Bethe lattice cannot capture consequences of a geometric spin frustration inherent to the antiferromagnetic Ising model on close-packed planar lattices, because the Bethe lattices do not contain any (odd-numbered) cycles. This deficiency can be avoided by considering the Ising model on another type of recursive lattice built up from the inter-connected polygons, whereas the interior part deep inside of such recursively built tree is usually referred to as the Husimi lattice  \cite{Monroe92,Monroe98,Ananikian98,Ananikian10,Jurcisin12,Bobak13,Bobak14,Huang14}. It is worthy to notice that the Ising and Ising-like models defined on the Husimi lattices have played an important role in a description of several interesting physical systems and phenomena like binary alloys, rare gases, lipid bilayers, spin liquids, spin glasses  or polymers (see Ref. \cite{Jurcisin12} and references cited therein).  

Recently, the spin-1/2 Ising-Heisenberg model on triangulated (triangles-in-triangles) planar lattices have attracted a great deal of attention \cite{Strecka08,Yao,Cis13a,Cis13b}, because it closely resembles a magnetic structure of real polymeric coordination compounds Cu$_9$X$_2$(cpa)$_6$$\cdot$nH$_2$O (X=F,Cl,Br) with the geometry of triangulated kagom\'e lattice \cite{norm87,norm90,gonz93}. Besides a clear experimental motivation, the spin-1/2 Ising-Heisenberg model on triangulated planar lattices has afforded a valuable playground for an investigation of the role of local quantum fluctuations within a novel class of exactly solvable classical-quantum spin systems. It has been convincingly evidenced \cite{Cis13a} that the spin-1/2 Ising-Heisenberg model with a greater connectivity of triangles-in-triangles units in the underlying triangular lattice gives rise to stronger local quantum fluctuations, which preferably support a remarkable quantum order at low enough temperatures instead of a quantum disorder through the order-by-disorder effect. Contrary to this, the spin-1/2 Ising-Heisenberg model on triangulated kagom\'e lattice cannot exhibit this intriguing feature \cite{Strecka08,Yao}. Bearing this in mind, it appears worthwhile to examine the spin-1/2 Ising-Heisenberg model on related triangulated Husimi lattices, in which the connectivity of triangles-in-triangles units can be varied systematically.   

The outline of this paper is as follows. The spin-1/2 Ising-Heisenberg model on triangulated Husimi lattices will be introduced in Sec.~\ref{sec:2} together with a few basic steps of its exact treatment. The most interesting results for the ground-state and finite-temperature phase diagrams will be discussed along with typical thermal variations of spontaneous magnetization in Sec.~\ref{sec:3}. The paper ends up with a brief summary of the most important findings in Sec.~\ref{sec:4}.

\begin{figure*}
\begin{center}
\includegraphics[width=10cm]{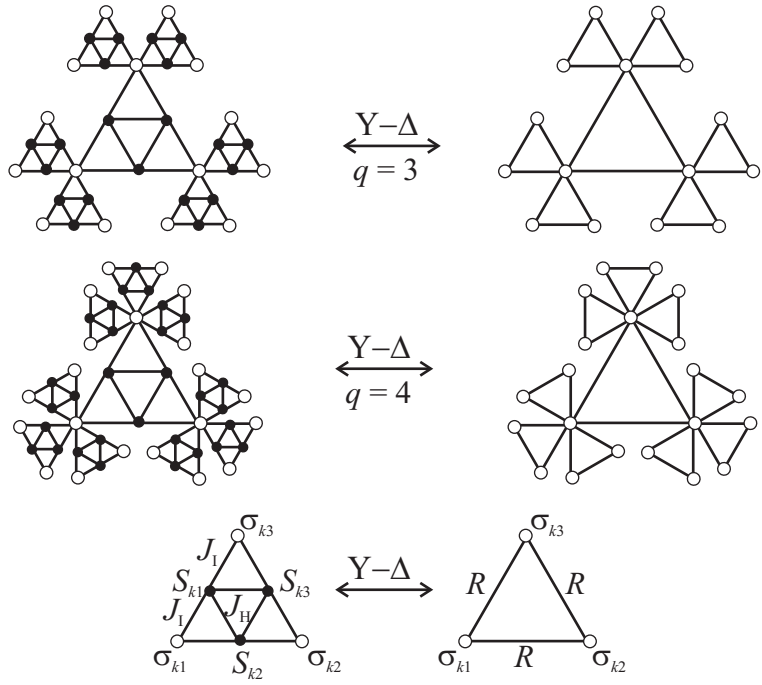} 
\end{center} 
\vspace{-4mm} 
\caption{The spin-1/2 Ising-Heisenberg model on the triangulated Husimi lattices (figures on the left) and its rigorous mapping equivalence with the effective spin-1/2 Ising model on the pure Husimi lattices (figures on the right) established by means of the generalized star-triangle transformation (figure at the bottom). The triangulated Husimi lattices, which include two different lattice sites occupied by the Ising (empty circles) and Heisenberg (full circles) spins, are drawn for two particular values of the coordination number $q=3$ and $4$.}
\label{fig1}
\end{figure*}

\section{Model and method}
\label{sec:2}
Let us consider the spin-1/2 Ising-Heisenberg model on triangulated Husimi lattices, which are schematically illustrated on the left-hand-side of Fig. \ref{fig1} for two selected values of the coordination number $q=3$ and $4$. The overall magnetic structure of triangulated Husimi lattices form smaller triangles of three quantum Heisenberg spins (Heisenberg trimers), which are embedded into each larger triangle of the pure Husimi lattice involving in its nodal lattice sites the classical Ising spins. The total Hamiltonian of the spin-1/2 Ising-Heisenberg model on triangulated Husimi lattice
denoting a deep interior such recursively built tree can be defined as follows
\begin{eqnarray}
\hat{\cal H} = -J_{\rm H} \sum_{\langle k,l \rangle}^{Nq} [\Delta(\hat{S}_{k}^{x}\hat{S}_{l}^{x}+\hat{S}_{k}^{y}\hat{S}_{l}^{y})+\hat{S}_{k}^{z}\hat{S}_{l}^{z}]
           -J_{\rm I} \sum_{\langle k,j \rangle}^{2Nq} \hat{S}_{k}^{z} \hat{\sigma}_{j}^{z}. \nonumber \\
\label{eq:1}
\end{eqnarray}
Here, $\hat{\sigma}^{z}_{j}$ and $\hat{S}_{k}^{\alpha}$ ($\alpha = x,y,z$) label spatial components of the usual spin-1/2 operator, the parameter $J_{\rm H}$ accounts for the XXZ Heisenberg interaction between the nearest-neighbor Heisenberg spins, $\Delta$ is a spatial anisotropy in this interaction term, the parameter $J_{\rm I}$ stands for the Ising interaction between the nearest-neighbor Heisenberg and Ising spins, respectively, $N$ denotes the total number of the Ising spins and $q$ is the coordination number that determines how many triangles-in-triangles units meet at each nodal site of the underlying Husimi lattice. The total Hamiltonian (\ref{eq:1}) of the spin-1/2 Ising-Heisenberg model on triangulated Husimi lattices alternatively can be rewritten as a sum over the cluster Hamiltonians
\begin{eqnarray}
\hat{\cal H}=\sum_{k=1}^{Nq/3} \hat{\cal H}_{k},
\label{eq:2}
\end{eqnarray}	
whereas each cluster Hamiltonian $\hat{\cal H}_{k}$ involves all the interaction terms of three Heisenberg spins from the $k$th triangle-in-triangle unit (i.e. the six-spin cluster schematically illustrated at the bottom of Fig. \ref{fig1}) 
\begin{eqnarray}
\hat{\cal H}_{k} = \!\!\! &-& \!\!\! J_{\rm H}\sum_{i=1}^{3}[\Delta (\hat S_{k,i}^x \hat S_{k,i+1}^x 
   + \hat S_{k,i}^y \hat S_{k,i+1}^y) + \hat S_{k,i}^z \hat S_{k,i+1}^z ] \nonumber \\
   \!\!\! &-& \!\!\! J_{\rm I}\sum_{i=1}^{3}\hat \sigma_{k,i}^z (\hat S_{k,i}^z + \hat S_{k,i+1}^z). \qquad (S_{k,4} \equiv S_{k,1})
\label{eq:3}
\end{eqnarray}
It is quite evident that the standard commutation relations $[\hat{\cal H}_i, \hat{\cal H}_j] = 0$ are satisfied for different cluster Hamiltonians $i\neq j$ and hence, the partition function of the spin-1/2 Ising-Heisenberg model on triangulated Husimi lattices can be partially factorized into the following product 
\begin{eqnarray}
{\cal Z}_{\rm IHM} = \sum_{\{ \sigma_i \}} \prod_{k = 1}^{Nq/3} \mbox{Tr}_k \exp(- \beta \hat {\cal H}_k)
               = \sum_{\{ \sigma_i \}} \prod_{k = 1}^{Nq/3} {\cal Z}_{k},
\label{eq:4}
\end{eqnarray}
where $\beta=1/(k_{\rm B}T)$, $k_{\rm B}$ is the Boltzmann's constant, $T$ is the absolute temperature, the symbol $\sum_{\{\sigma_{i}\}}$ denotes a summation over spin states of all the Ising spins and the symbol $\mbox{Tr}_{k} = \mbox{Tr}_{S_{k1}} \mbox{Tr}_{S_{k2}} \mbox{Tr}_{S_{k3}}$ stands for a trace over spin degrees of freedom of the $k$th Heisenberg trimer. The mathematical structure of Eq. (\ref{eq:4}) implies that one may perform a trace over spin degrees of freedom of different Heisenberg trimers independently of each other in order to obtain an explicit form of the Boltzmann's weight ${\cal Z}_{k}$, which will solely depend on three nodal Ising spins $\sigma_{k1}, \sigma_{k2}$, and $\sigma_{k3}$ attached to the $k$th Heisenberg trimer. Afterwards, one may take advantage of the generalized star-triangle transformation \cite{Fis59,Syo72,Roj09,Str10}
 \begin{eqnarray}
{\cal Z}_k \!\!\! && \!\!\! (\sigma_{k1}^z, \sigma_{k2}^z,\sigma_{k3}^z) = \mbox{Tr}_k \exp(- \beta \hat {\cal H}_k) \nonumber \\
\!\!\!\! && \!\!\!\!=A \exp[\beta R (\sigma_{k1}^z \sigma_{k2}^z+\sigma_{k2}^z \sigma_{k3}^z+\sigma_{k3}^z \sigma_{k1}^z)], 
\label{eq:5}
\end{eqnarray}
which replaces the effective Boltzmann's weight ${\cal Z}_k$ with the equivalent expression depending on three nodal Ising spins only. The star-triangle mapping transformation generally represents a set of eight algebraic equations, which can be obtained from Eq. (\ref{eq:5}) by substituting all available spin configurations of three Ising spins involved therein. In fact, one gets just two independent equations; the first equation for two uniform configurations with the three equally aligned Ising spins
\begin{eqnarray}
Z_1 &=& 2 \exp \left(\frac{3}{4}\beta J_{\rm H}\right) \cosh\left(\frac{3}{2}\beta J_{\rm I}\right) \nonumber \\
&+& 2 \exp\left[-\frac{1}{4}\beta J_{\rm H} \left(1 - 4 \Delta\right)\right]\cosh \left(\frac{1}{2}\beta J_{\rm I}\right)\nonumber \\
&+& 4 \exp\left[-\frac{1}{4}\beta J_{\rm H} \left(1 + 2 \Delta\right)\right] \cosh \left(\frac{1}{2}\beta J_{\rm I}\right),\nonumber \\
&=& A\exp\left(\frac{3}{4}\beta R\right),
\label{eq:6}
\end{eqnarray}
and the second one for six nonuniform configurations with one Ising spin pointing in opposite with respect to the other two
\begin{eqnarray}
Z_2 &=& 2 \exp \left(\frac{3}{4}\beta J_{\rm H}\right) \cosh\left(\frac{1}{2}\beta J_{\rm I}\right)\nonumber \\ 
&+& 2 \exp\left[-\frac{1}{4}\beta J_{\rm H} \left(1+2\Delta\right)\right]\cosh\left(\frac{1}{2}\beta J_{\rm I}\right)\nonumber \\ 
&+& 2 \exp\left[-\frac{1}{4}\beta J_{\rm H} \left(1-\Delta\right)\right] \cosh \left(\frac{1}{2}\beta Q^+\right)\nonumber \\ 
&+& 2 \exp\left[-\frac{1}{4}\beta J_{\rm H} \left(1-\Delta\right)\right] \cosh\left (\frac{1}{2}\beta Q^-\right),\nonumber \\ 
&=& A\exp\left(-\frac{1}{4}\beta R\right),
\label{eq:7}
\end{eqnarray}
which involves the parameters $Q^\pm$ defined as follows
\begin{eqnarray}
Q^\pm= \sqrt{\left(\frac{J_{\rm H}\Delta}{2}\pm J_{\rm I}\right)^2+2\left(J_{\rm H}\Delta\right)^2}. 
\label{eq:8}
\end{eqnarray}
It is noteworthy that the algebraic equations (\ref{eq:6}) and (\ref{eq:7}) unambiguously determine yet unspecified mapping parameters $A$ and $R$ 
\begin{eqnarray}
A = \left(Z_{1}Z_{2}^3\right)^{1/4}, \qquad \beta R = \ln \left(\frac{Z_1}{Z_2}\right), 
\label{eq:9}
\end{eqnarray}
which enter into the generalized star-triangle transformation (\ref{eq:5}). On assumption that the mapping parameters $A$ and $R$ satisfy Eqs. (\ref{eq:6})-(\ref{eq:9}), the star-triangle transformation (\ref{eq:5}) holds quite generally (regardless of spin states of three nodal Ising spins) and it can be substituted into the relation (\ref{eq:4}) with the aim to get a rigorous mapping correspondence between the partition function ${\cal Z}_{\rm IHM}$ of the spin-1/2 Ising-Heisenberg model on the triangulated Husimi lattices and, respectively, the partition function ${\cal Z}_{\rm IM}$ of the effective spin-1/2 Ising model on a pure Husimi lattice  
\begin{equation}
{\cal Z}_{\rm IHM} (\beta, J_{\rm H}, J_{\rm I}, \Delta) = A^{Nq/3} {\cal Z}_{\rm IM} (\beta, R),
\label{eq:10}
\end{equation}
which is unambiguously given by the Hamiltonian with the effective nearest-neighbor Ising interaction $R$: 
\begin{eqnarray}
{\cal H}_{\rm IM} = -R\sum_{\langle i,j \rangle }^{Nq} \sigma_{i}^z \sigma_{j}^z.
\label{eq:11}
\end{eqnarray}

It should be noted that the effective spin-1/2 Ising model on pure Husimi lattices can be exactly solved by means of the method of exact recursion relations, which consequently enables to extract exact results for the spin-1/2 Ising-Heisenberg model on the triangulated Husimi lattices from the relevant mapping correspondence between both models. The rigorous treatment of the spin-1/2 Ising model on triangular Husimi lattices using the recursion method has been reported on previously in several foregoing works \cite{Monroe92,Monroe98,Ananikian98,Ananikian10,Jurcisin12,Bobak13,Bobak14}, so it is sufficient to recall the most important outcomes of this calculation. It can be easily proved that the effective nearest-neighbor coupling (\ref{eq:9}) of the corresponding spin-1/2 Ising model on pure Husimi lattices is always ferromagnetic (i.e. $R>0$), which substantially simplifies the iteration procedure as three inter-connected recursion relations can be replaced with a single recursion relation. Assuming the ferromagnetic pair interaction $R>0$ between the nearest-neighbor spins, the spontaneous magnetization of the spin-1/2 Ising model on triangular Husimi lattices with the coordination number $q$ 
can be expressed as 
\begin{eqnarray}
m_{\rm IM} \equiv \langle \sigma_{k1}^z \rangle_{\rm IM} =  \frac{1}{2} \left(\frac{1-x^q}{1+x^q}\right),
\label{eq:13} 
\end{eqnarray}
where the symbol $\langle \cdots \rangle_{\rm IM}$ denotes a canonical ensemble average performed within the effective Ising model defined by the Hamiltonian (\ref{eq:11}) and $x$ represents the stable fixed point of the recurrence relation (i.e. $x = \lim_{n \to \infty} x_n$) 
\begin{eqnarray}
x_{n} = \frac{1+2x_{n-1}^{q-1}+\exp(\beta R)x_{n-1}^{2q-2}}{\exp(\beta R)+2x_{n-1}^{q-1}+x_{n-1}^{2q-2}}.
\label{eq:14}
\end{eqnarray} 
Similarly, the expectation values of other important quantities such as the triplet correlation function between three Ising spins residing sites of an elementary triangular plaquette of the pure Husimi lattice can also be calculated from the stable fixed point of the recurrence relation (\ref{eq:14}) according to the formula 
\begin{eqnarray}
t_{\rm IM} \equiv \left \langle \sigma_{k1}^z \sigma_{k2}^z \sigma_{k3}^z \right \rangle_{\rm IM} = \frac{1}{8} \left(\frac{y-zx^{q-1}}{1+x^q}\right),
\label{eq:18}
\end{eqnarray}
which involves another two parameters $y$ and $z$ defined through the stable fixed point of recurrence relation (\ref{eq:14}) 
\begin{eqnarray}
y &=& \frac{\exp(\beta R)-2x^{q-1}+x^{2q-2}}{\exp(\beta R)+2x^{q-1}+x^{2q-2}}, \nonumber \\
z &=& \frac{1-2x^{q-1}+\exp(\beta R)x^{2q-2}}{\exp(\beta R)+2x^{q-1}+x^{2q-2}}.
\label{eq:19}
\end{eqnarray}		
It has been rigorously proved by Jur\v{c}i\v{s}inov\'a and Jur\v{c}i\v{s}in \cite{Jurcisin12} that the spin-1/2 Ising model on triangular Husimi lattices with the ferromagnetic nearest-neighbor interaction $R>0$ exhibits a continuous (second-order) phase transition from a ferromagnetic phase to a paramagnetic phase at the following (inverse) critical temperature 
\begin{eqnarray}
\beta_{\rm c} R = \frac{R}{k_{\rm B} T_{\rm c}} = \ln \left( \frac{2q+1}{2q-3} \right). 
\label{eq:cc}
\end{eqnarray} 

With all this in mind, let us extract exact results for the spin-1/2 Ising-Heisenberg model on the triangulated Husimi lattices from the corresponding rigorous results of the effective spin-1/2 Ising model on triangular Husimi lattices. It can be straightforwardly proved by adopting the exact mapping theorems developed by Barry et al. \cite{barr88,khat90,barr91,barr95} that the sublattice magnetization of the Ising spins within the spin-1/2 Ising-Heisenberg model on triangulated Husimi lattices directly equals to the spontaneous magnetization of the effective spin-1/2 Ising model on pure Husimi lattices   
\begin{eqnarray}
m_{\rm I} \equiv \langle \hat{\sigma}_{k1}^z \rangle_{\rm IHM} = \langle \sigma_{k1}^z \rangle_{\rm IM} \equiv m_{\rm IM} (\beta R).
\label{eq:12} 
\end{eqnarray}
Here, the symbol $\langle \cdots \rangle_{\rm IHM}$ denotes a canonical ensemble average performed within the spin-1/2 Ising-Heisenberg model on the triangulated Husimi lattice defined through the Hamiltonian (\ref{eq:1}) and the single-site magnetization of the effective spin-1/2 Ising model on the triangular Husimi lattices can be calculated from Eqs. (\ref{eq:13}) and (\ref{eq:14}) after solving the latter recursion relation iteratively. Contrary to this, the sublattice magnetization of the Heisenberg spins in the spin-1/2 Ising-Heisenberg model on triangulated Husimi lattices can be exactly calculated from the generalized form of Callen-Suzuki spin identity \cite{Callen,Suzuki,balc02}
\begin{eqnarray}
m_{\rm H} = \left \langle \hat{S}_{k1}^z \right \rangle_{\rm IHM} =  \left \langle \frac{\mbox{Tr}_k \hat{S}_{k1}^z \exp(-\beta \hat{\cal H}_k)}{\mbox{Tr}_k \exp(- \beta \hat {\cal H}_k)}\right \rangle_{\rm IHM}.
\label{eq:15}
\end{eqnarray}
According to Eq. (\ref{eq:15}), the sublattice magnetization of the Heisenberg spins $m_{\rm H}$ can be related to the formerly derived single-site magnetization of the Ising spins $m_{\rm I}$ given by Eqs. (\ref{eq:13})-(\ref{eq:14}) and the triplet correlation function $t_{\rm IM}$ given by Eqs. (\ref{eq:18})-(\ref{eq:19})  
\begin{eqnarray}
m_{\rm H}=\frac{m_{\rm IM}}{2} \left [\frac{Q_{1}}{Z_{1}} +\frac{Q_{2}}{Z_{2}} \right ]+\frac{2t_{\rm IM} }{3} \left [\frac{Q_{1}}{Z_{1}} -3\frac{Q_{2}}{Z_{2}}\right].
\label{eq:16}
\end{eqnarray}
Two new functions $Q_1$ and $Q_2$ entering the exact formula (\ref{eq:16}) for the sublattice magnetization of the Heisenberg spins are defined as follows 
\begin{eqnarray}
Q_1&=&3 \exp \left(\frac{3}{4}\beta J_{\rm H}\right) \cosh\left(\frac{3}{2}\beta J_{\rm I}\right)\nonumber \\
&+& \exp\left[-\frac{1}{4}\beta J_{\rm H} \left(1 - 4 \Delta\right)\right]\sinh \left(\frac{1}{2}\beta J_{\rm I}\right)\nonumber \\
&+& 2 \exp\left[-\frac{1}{4}\beta J_{\rm H} \left(1 + 2 \Delta\right)\right] \sinh \left(\frac{1}{2}\beta J_{\rm I}\right),\nonumber \\
Q_2&=&3 \exp \left(\frac{3}{4}\beta J_{\rm H}\right) \sinh\left(\frac{1}{2}\beta J_{\rm I}\right)\nonumber \\
&+& \exp\left[-\frac{1}{4}\beta J_{\rm H} \left(1 + 2 \Delta\right)\right]\sinh \left(\frac{1}{2}\beta J_{\rm I}\right)\nonumber \\
&+& \exp\left[-\frac{1}{4}\beta J_{\rm H} \left(1 -\Delta\right)\right] \cosh \left(\frac{1}{2}\beta Q^+\right),\nonumber \\
&-& \exp\left[-\frac{1}{4}\beta J_{\rm H} \left(1 -  \Delta\right)\right]\cosh \left(\frac{1}{2}\beta Q^-\right).
\label{eq:17}
\end{eqnarray}
Last but not least, the critical points of the spin-1/2 Ising-Heisenberg model on triangulated Husimi lattices can be readily obtained from a comparison of the effective coupling (\ref{eq:9}) with its critical value (\ref{eq:cc}), which subsequently provides the following explicit form for the critical condition 
\begin{eqnarray}
(2q-3) Z_1^{\rm c} = (2q+1) Z_2^{\rm c},
\label{eq:tc}
\end{eqnarray}
where the superscript c means that the inverse critical temperature $\beta_{\rm c}$ enters into the expressions $Z_1$ and $Z_2$ defined by Eqs. (\ref{eq:6}) and (\ref{eq:7}) instead of the usual inverse temperature $\beta$. The critical condition (\ref{eq:tc}) can be rather straightforwardly used for constructing finite-temperature phase diagrams of the spin-1/2 Ising-Heisenberg model on triangulated Husimi lattices with arbitrary value of the coordination number $q$.

\section{Results and discussion}
\label{sec:3}
Let us proceed to a discussion of the most interesting results for the spin-1/2 Ising-Heisenberg model on triangulated Husimi lattices. In the following, we will be mainly interested in addressing the question of whether the coordination number of the underlying Husimi lattice may essentially influence the overall magnetic behavior of the studied model or not. In general, the change of the ferromagnetic Ising interaction ($J_{\rm I}>0$) to the antiferromagnetic one ($J_{\rm I}<0$) does not cause any fundamental difference in a respective magnetic behavior and hence, our further attention will be merely restricted to a special case with the ferromagnetic Ising coupling $J_{\rm I}>0$ henceforth serving as the energy unit. By contrast, the antiferromagnetic Heisenberg interaction $J_{\rm H}<0$ is responsible for a mutual interplay of local quantum fluctuations with geometric spin frustration totally absent in the model with the ferromagnetic Heisenberg interaction $J_{\rm H}>0$, which makes an investigation of the highly frustrated region $J_{\rm H}/J_{\rm I} \ll 0$ particularly intriguing. 

\begin{figure}[t] 
\vspace{0cm}
\includegraphics[width=8cm]{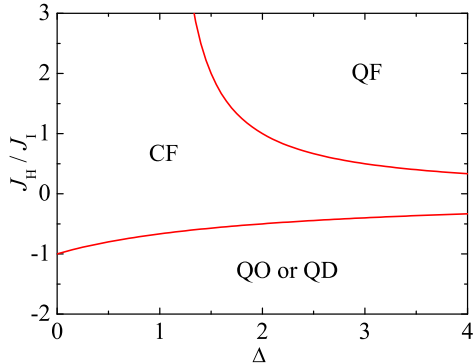}
\vspace{-0.5cm}
\caption{\small (Color online) The ground-state phase-diagram of the spin-1/2 Ising-Heisenberg model on the triangulated Husimi lattices in the $\Delta-J_{\rm H}/J_{\rm I}$ plane.}
\label{fig2}
\end{figure}

Let us examine first a ground state of the spin-1/2 Ising-Heisenberg model on triangulated Husimi lattices. It is worth mentioning that the overall ground-state eigenvector is given by a tensor product over the lowest-energy eigenstates of the cluster Hamiltonians (\ref{eq:3}), which means that the ground-state phase diagram and individual ground states of the spin-1/2 Ising-Heisenberg model on triangulated Husimi lattices are completely identical with that reported on previously for the analogous model on triangulated planar lattices composed of the same structural unit \cite{Strecka08,Yao,Cis13a}. In this regard, let us merely quote individual ground-state eigenvectors of the spin-1/2 Ising-Heisenberg model on triangulated Husimi lattices without their detailed description (the interested reader is referred to Ref. \cite{Cis13a} for further details). Altogether, the ground-state phase diagram depicted in Fig. \ref{fig2} in the $\Delta-J_{\rm H}/J_{\rm I}$ plane includes two unique spontaneously ordered ground states, namely, the classical ferromagnetic (CF) ground state 
\begin{eqnarray}
\vert \mbox{CF} \rangle = \prod_{i=1}^{N} \left  \vert \uparrow \right \rangle_{\!\sigma_{i}^z} 
                           \prod_{k=1}^{Nq/3} \left \vert \uparrow \uparrow \uparrow \right \rangle_{\!S_{k1}^z,S_{k2}^z,S_{k3}^z},
\label{cfp} 
\end{eqnarray}
the quantum ferromagnetic (QF) ground state
\begin{eqnarray}
\vert \mbox{QF} \rangle \!=\! \prod_{i=1}^{N} \! \left  \vert \uparrow \right \rangle_{\!\sigma_{i}^z} \!\! \prod_{k=1}^{Nq/3} \!\! \frac{1}{\sqrt{3}} \! 
\left( \left \vert \uparrow \uparrow \downarrow \right \rangle \!+\! \left \vert \uparrow \downarrow \uparrow \right \rangle 
\!+\! \left \vert \downarrow \uparrow \uparrow \right \rangle \right)_{\!S_{k1}^z,S_{k2}^z,S_{k3}^z}\!\!\!,
\label{qfp} 
\end{eqnarray} 
and one highly degenerate ground-state manifold spanned over the following 10 eigenstates of the cluster Hamiltonians (\ref{eq:3}) 
\begin{eqnarray}
\begin{array}{llll} 
\left  \vert \uparrow \uparrow \uparrow \right \rangle_{\!\sigma_{k1}^z,\sigma_{k2}^z,\sigma_{k3}^z} \frac{1}{\sqrt{2}} \!
\left( \left \vert \uparrow \downarrow \uparrow \right \rangle \!-\! \left \vert \downarrow \uparrow \uparrow \right \rangle \right)_{S_{k1}^z, S_{k2}^z, S_{k3}^z} \\
\left  \vert \uparrow \uparrow \uparrow \right \rangle_{\!\sigma_{k1}^z,\sigma_{k2}^z,\sigma_{k3}^z} \frac{1}{\sqrt{2}} \!
\left( \left \vert \uparrow \uparrow \downarrow \right \rangle \!-\! \left \vert \downarrow \uparrow \uparrow \right \rangle \right)_{S_{k1}^z, S_{k2}^z, S_{k3}^z}  \\
\left  \vert \downarrow \downarrow \downarrow \right \rangle_{\!\sigma_{k1}^z,\sigma_{k2}^z,\sigma_{k3}^z} \frac{1}{\sqrt{2}} \!
\left( \left \vert \uparrow \downarrow \downarrow \right \rangle \!-\! \left \vert \downarrow \uparrow \downarrow \right \rangle \right)_{S_{k1}^z, S_{k2}^z, S_{k3}^z} \\
\left  \vert \downarrow \downarrow \downarrow \right \rangle_{\!\sigma_{k1}^z,\sigma_{k2}^z,\sigma_{k3}^z} \frac{1}{\sqrt{2}} \!
\left( \left \vert \uparrow \downarrow \downarrow \right \rangle \!-\! \left \vert \downarrow \downarrow \uparrow \right \rangle \right)_{S_{k1}^z, S_{k2}^z, S_{k3}^z}  \\
\left  \vert \downarrow \uparrow \uparrow \right \rangle_{\!\sigma_{k1}^z,\sigma_{k2}^z,\sigma_{k3}^z} \frac{1}{\sqrt{2}} \!
\left( \left \vert \uparrow \downarrow \uparrow \right \rangle \!-\! \left \vert \downarrow \uparrow \uparrow \right \rangle \right)_{S_{k1}^z, S_{k2}^z, S_{k3}^z} \\
\left  \vert \uparrow \downarrow \uparrow \right \rangle_{\!\sigma_{k1}^z,\sigma_{k2}^z,\sigma_{k3}^z} \frac{1}{\sqrt{2}} \!
\left( \left \vert \uparrow \uparrow \downarrow \right \rangle \!-\! \left \vert \downarrow \uparrow \uparrow \right \rangle \right)_{S_{k1}^z, S_{k2}^z, S_{k3}^z} \\
\left  \vert \uparrow \uparrow \downarrow \right \rangle_{\!\sigma_{k1}^z,\sigma_{k2}^z,\sigma_{k3}^z} \frac{1}{\sqrt{2}} \!
\left( \left \vert \uparrow \uparrow \downarrow \right \rangle \!-\! \left \vert \uparrow \downarrow \uparrow \right \rangle \right)_{S_{k1}^z, S_{k2}^z, S_{k3}^z} \\
\left  \vert \uparrow \downarrow \downarrow \right \rangle_{\!\sigma_{k1}^z,\sigma_{k2}^z,\sigma_{k3}^z} \frac{1}{\sqrt{2}} \!
\left( \left \vert \uparrow \downarrow \downarrow \right \rangle \!-\! \left \vert \downarrow \uparrow \downarrow \right \rangle \right)_{S_{k1}^z, S_{k2}^z, S_{k3}^z} \\
\left  \vert \downarrow \uparrow \downarrow \right \rangle_{\!\sigma_{k1}^z,\sigma_{k2}^z,\sigma_{k3}^z} \frac{1}{\sqrt{2}} \!
\left( \left \vert \uparrow \downarrow \downarrow \right \rangle \!-\! \left \vert \downarrow \downarrow \uparrow \right \rangle \right)_{S_{k1}^z, S_{k2}^z, S_{k3}^z} \\
\left  \vert \downarrow \downarrow \uparrow \right \rangle_{\!\sigma_{k1}^z,\sigma_{k2}^z,\sigma_{k3}^z} \frac{1}{\sqrt{2}} \!
\left( \left \vert \downarrow \uparrow \downarrow \right \rangle \!-\! \left \vert \downarrow \downarrow \uparrow \right \rangle \right)_{S_{k1}^z, S_{k2}^z, S_{k3}^z}. 
        \end{array}  
\label{odp}         
\end{eqnarray}
It should be noted that the highly degenerate ground-state manifold (\ref{odp}) basically changes its character in the limiting Ising case $\Delta = 0$, which does not exhibit any interesting feature due to a complete lack of local quantum fluctuations and this special case will be therefore left out from our subsequent ground-state considerations. Although the spin-1/2 Ising-Heisenberg model on most triangulated planar lattices \cite{Strecka08,Yao,Cis13a} prefers a quantum disorder (QD) in a highly frustrated parameter region pertinent to the ground-state manifold (\ref{odp}), it has been convincingly evidenced in Ref. \cite{Cis13a} that the model alternatively may display a spectacular spontaneous quantum order (QO) due to the  order-by-disorder effect. These results would suggest that the geometry of underlying magnetic lattice can basically influence an essence of the ground-state manifold (\ref{odp}), which is usually disordered but under certain conditions may become spontaneously ordered. 

Let us find a rigorous criterion that will definitely resolve the puzzling issue concerning with the spontaneously ordered or disordered nature of the ground-state manifold (\ref{odp}) of the spin-1/2 Ising-Heisenberg model on triangulated Husimi lattices. It is quite evident from Eqs. (\ref{eq:12}) and (\ref{eq:16}) that a nonzero spontaneous magnetization of the effective spin-1/2 Ising model on the pure Husimi lattices represents sought condition ensuring nonzero spontaneous magnetizations of the Ising and Heisenberg spins. The spin-1/2 Ising model on the triangular Husimi lattices has nonzero spontaneous magnetization if and only if the effective nearest-neighbor coupling $\beta R$ exceeds the critical value given by Eq. (\ref{eq:cc}), which monotonically decreases with rising coordination number of the underlying Husimi lattice (e.g. $\beta_{\rm c} R = \ln (5)$ for $q=2$, $\beta_{\rm c} R = \ln (7/3)$ for $q=3$, $\beta_{\rm c} R = \ln (9/5)$ for $q=4$, etc.). In the highly frustrated region $J_{\rm H}/J_{\rm I} < - 2/(2 + \Delta)$ relevant to the ground-state manifold (\ref{odp}), the effective coupling acquires the zero-temperature asymptotic value $\lim_{T \to 0} \beta R = \ln 2$ that is conclusive whether or not a spontaneous long-range order appears. A comparison of the zero-temperature asymptotic limit $\lim_{T \to 0} \beta R = \ln 2$ with the critical value (\ref{eq:cc}) furnishes a rigorous proof that the ground-state manifold (\ref{odp}) remains disordered just for the spin-1/2 Ising-Heisenberg model on the triangulated Husimi lattices with the coordination numbers $q=2$ and $3$, while the same model on the triangulated Husimi lattices with greater values of the coordination number $q \geq 4$ shows a peculiar spontaneous long-range order of quantum nature.    

\begin{figure}[t] 
\vspace{0cm}
\includegraphics[width=8cm]{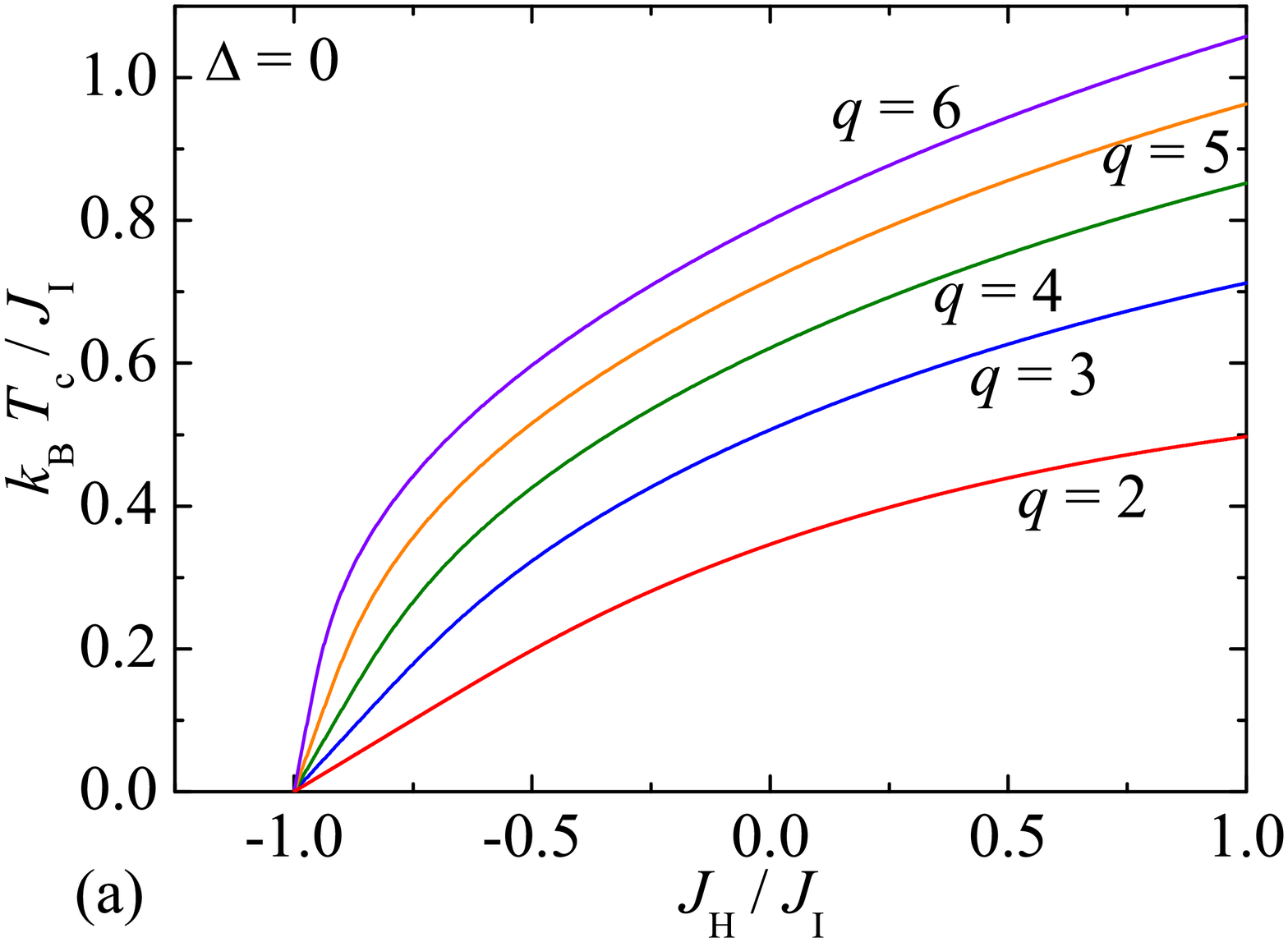}
\includegraphics[width=8cm]{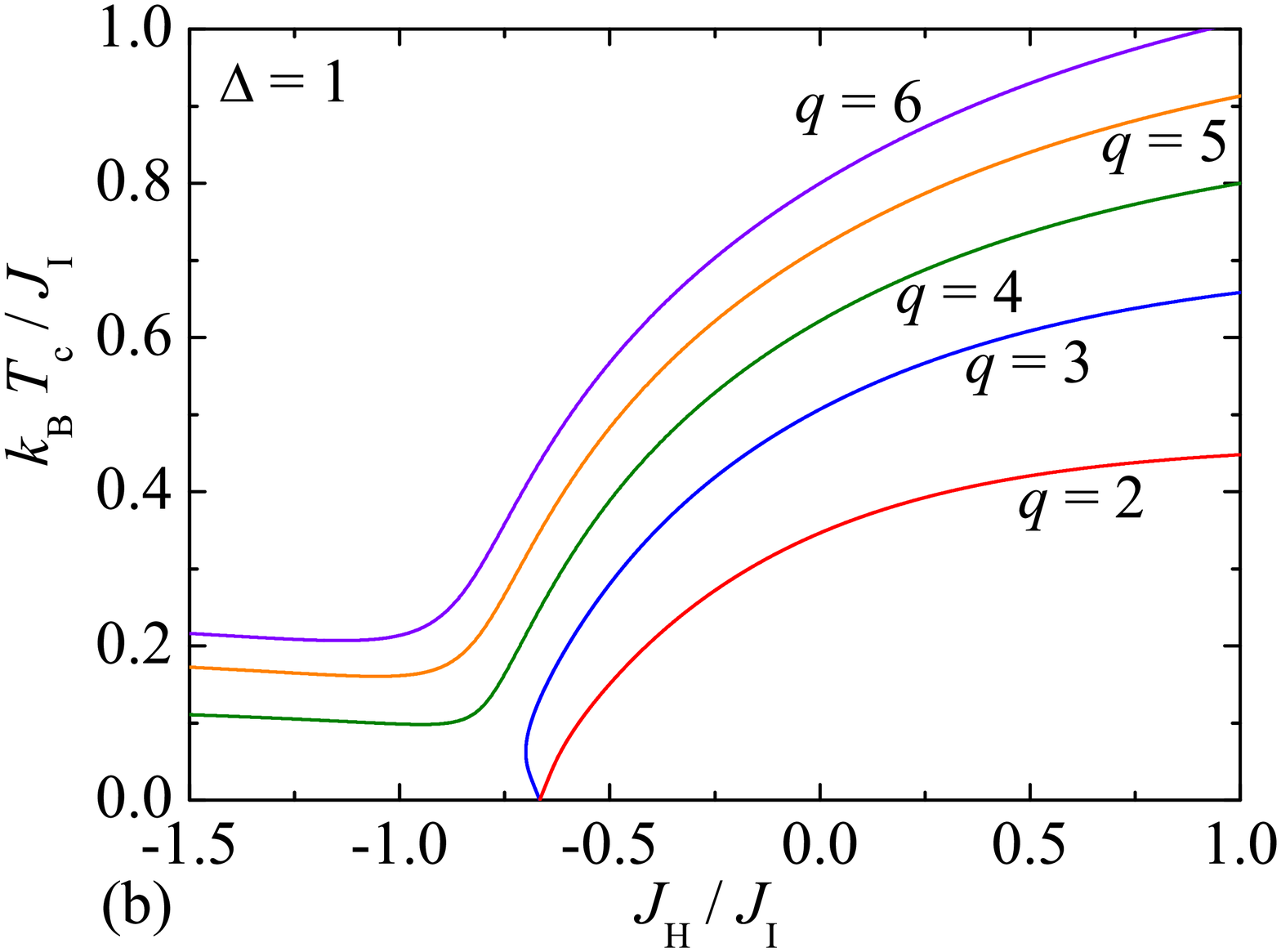}
\includegraphics[width=8cm]{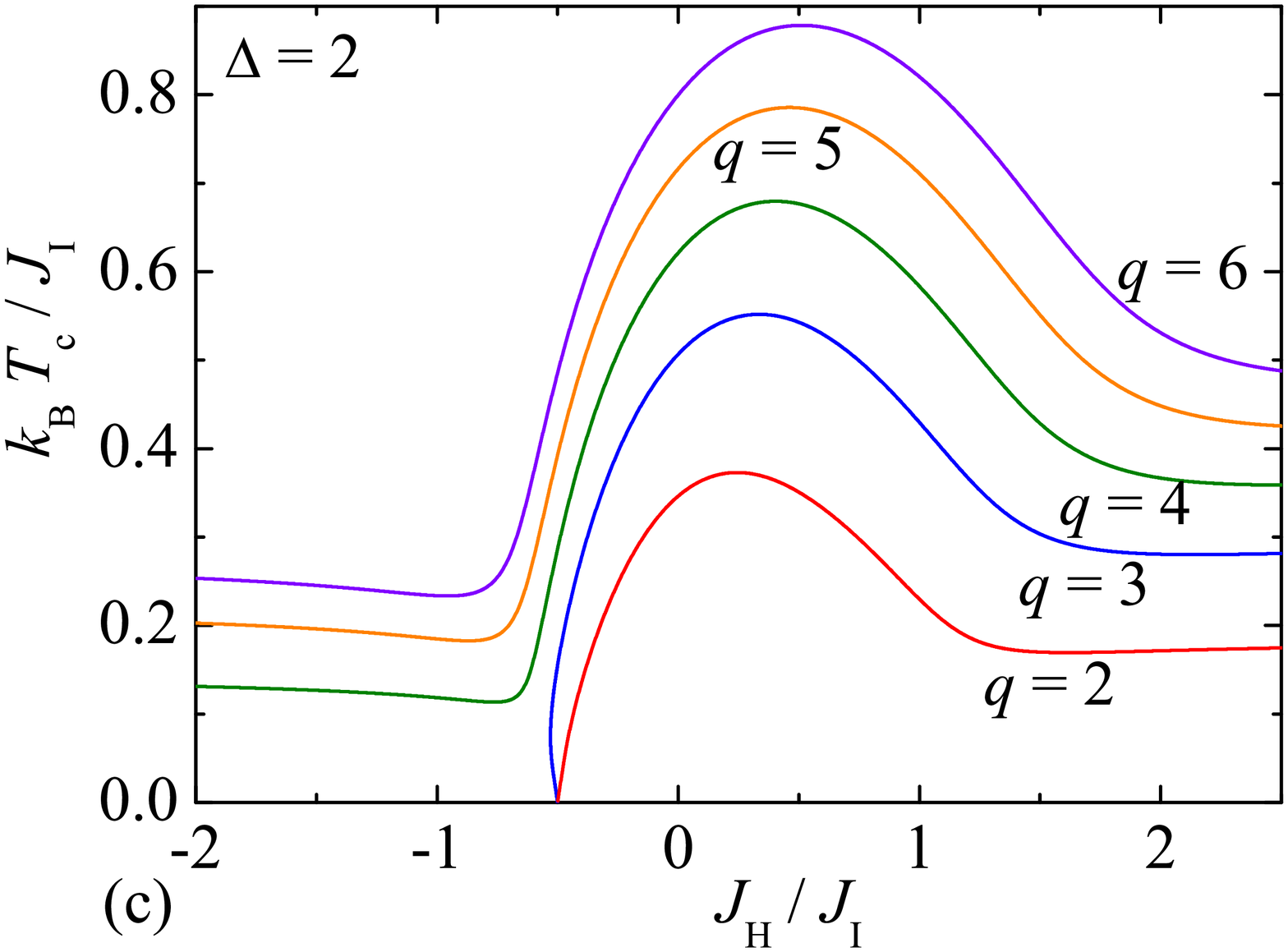}
\vspace*{-0.2cm}
\caption{\small (Color online) The critical temperature as a function of the ratio between the Heisenberg and Ising interaction for several values of the coordination number $q$ and three different values of the exchange anisotropy: (a) $\Delta = 0$; (b) $\Delta = 1$; and (c) $\Delta = 2$.}
\label{fig3}
\end{figure}

To verify our theoretical prediction of the quantum order-by-disorder phenomenon, let us explore in detail finite-temperature phase diagrams of the spin-1/2 Ising-Heisenberg model on the triangulated Husimi lattices, which are plotted in Fig. \ref{fig3} for three different values of the exchange anisotropy $\Delta$ and a few values of the coordination number $q$. First, it can be seen from Fig. \ref{fig3}(a) that the spin-1/2 Ising model on triangulated Husimi lattices retrieved in the limit $\Delta=0$ shows a familiar monotonous dependence of the critical temperature on the interaction ratio $J_{\rm H}/J_{\rm I}$, which becomes zero whenever the antiferromagnetic interaction $J_{\rm H}/J_{\rm I} \leq -1$ is strong enough to force the investigated model to the disordered ground state. The highest value of the critical temperature is accordingly obtained in the asymptotic limit $J_{\rm H}/J_{\rm I} \to \infty$
\begin{eqnarray}
\displaystyle \lim_{\frac{J_{\rm H}}{J_{\rm I}} \to \infty} \!\! \frac{k_{\rm B} T_{\rm c}}{J_{\rm I}} 
= \! \left \{ \ln \left[\! \frac{2q - 1 + 2 \sqrt{2(q - 1)}}{2q - 3} \right] \right\}^{-1}. 
\label{tcas1} 
\end{eqnarray} 
 
On the contrary, the spin-1/2 Ising-Heisenberg model on the triangulated Husimi lattices exhibits more diverse finite-temperature phase diagrams due to a mutual interplay between the geometric spin frustration and local quantum fluctuations, which are only present if the anisotropy parameter $\Delta>0$. Fig. \ref{fig3}(b) illustrates variations of the critical temperature with the interaction ratio $J_{\rm H}/J_{\rm I}$ that are quite typical for the model with the isotropic Heisenberg coupling $\Delta=1$ (the displayed case) as well as the XXZ Heisenberg coupling of the easy-axis type $0<\Delta<1$. Both aforementioned cases indeed show qualitatively the same dependences except a small quantitative difference including the asymptotic limit $J_{\rm H}/J_{\rm I} \to \infty$, at which the critical temperature still reaches the asymptotic value (\ref{tcas1}) in the case of easy-axis anisotropy $0<\Delta<1$ but it goes to a slightly lower asymptotic value in the isotropic case $\Delta=1$. However, the most important finding relates to the critical frontiers in a highly frustrated region. In accordance with our ground-state argumentation, the critical temperature tends to zero at the ground-state boundary ($J_{\rm H}/J_{\rm I} = -2/3$ for $\Delta=1$) between the classical ferromagnetic phase (\ref{cfp}) and the highly degenerate ground-state manifold (\ref{odp}) just for the spin-1/2 Ising-Heisenberg model on two triangulated Husimi lattices with the coordination numbers $q=2$ and $3$. Note furthermore that the model defined on the triangulated Husimi lattice with the coordination number $q=3$ displays a subtle reentrant region when the interaction ratio is selected slightly below the relevant ground-state phase boundary $J_{\rm H}/J_{\rm I} \lesssim -2/3$. On the other hand, the critical temperature of the spin-1/2 Ising-Heisenberg model on triangulated Husimi lattices with greater coordination numbers $q \geq 4$ remains nonzero in the whole frustrated region ($J_{\rm H}/J_{\rm I} \leq -2/3$ for $\Delta=1$), which provides an independent confirmation of the spectacular quantum order arising out from the highly degenerate ground-state manifold (\ref{odp}) due to the quantum order-by-disorder effect. The greater the coordination number of the triangulated Husimi lattice is, the higher is the critical temperature of the spontaneous quantum  long-range order.  

The last finite-temperature phase diagram shown in Fig. \ref{fig3}(c) illustrates the critical temperature of the spin-1/2 Ising-Heisenberg model on the triangulated Husimi lattices when assuming the XXZ Heisenberg coupling of the easy-plane type $\Delta>1$. Two appropriate comments are in order. First, it is quite apparent from Fig. \ref{fig3}(b) and Fig. \ref{fig3}(c) that the critical temperature qualitatively retains character of its dependence on the interaction ratio $J_{\rm H}/J_{\rm I}$ in the highly frustrated region for any $\Delta>0$, i.e. regardless of whether the Heisenberg interaction has the easy-axis anisotropy, the easy-plane anisotropy or is completely isotropic. Second, the critical temperature unexpectedly shows a considerable decrease upon strengthening of the ferromagnetic Heisenberg coupling in a nonfrustrated regime $J_{\rm H}/J_{\rm I}>0$. The intriguing suppression of the critical temperature can be attributed to a phase transition from the classical ferromagnetic phase (\ref{cfp}) to the quantum ferromagnetic phase (\ref{qfp}), which occurs at $J_{\rm H}/J_{\rm I}=1$ for the selected value of the exchange anisotropy $\Delta=2$. Owing to this fact, the critical temperature approaches the asymptotic limit 
\begin{eqnarray}
\displaystyle \lim_{\frac{J_{\rm H}}{J_{\rm I}} \to \infty} \!\! \frac{k_{\rm B} T_{\rm c}}{J_{\rm I}} 
= \frac{1}{3} \left\{\! \ln \left[\! \frac{2q - 1 + 2 \sqrt{2(q - 1)}}{2q - 3} \right] \right\}^{-1}, 
\label{tcas2} 
\end{eqnarray}
which is for the quantum ferromagnetic phase exactly three times lower than the one (\ref{tcas1}) of the classical ferromagnetic phase on account of a quantum reduction of the spontaneous magnetization of the Heisenberg spins to one-third of its saturation value. Hence, it follows that the quantum entanglement and associated quantum fluctuations are responsible for a gradual suppression of the critical temperature of the quantum ferromagnetic phase (\ref{qfp}) in response to a strengthening of the ferromagnetic Heisenberg interaction of easy-plane type. 
 
\begin{figure}[t] 
\vspace{0cm}
\includegraphics[width=8cm]{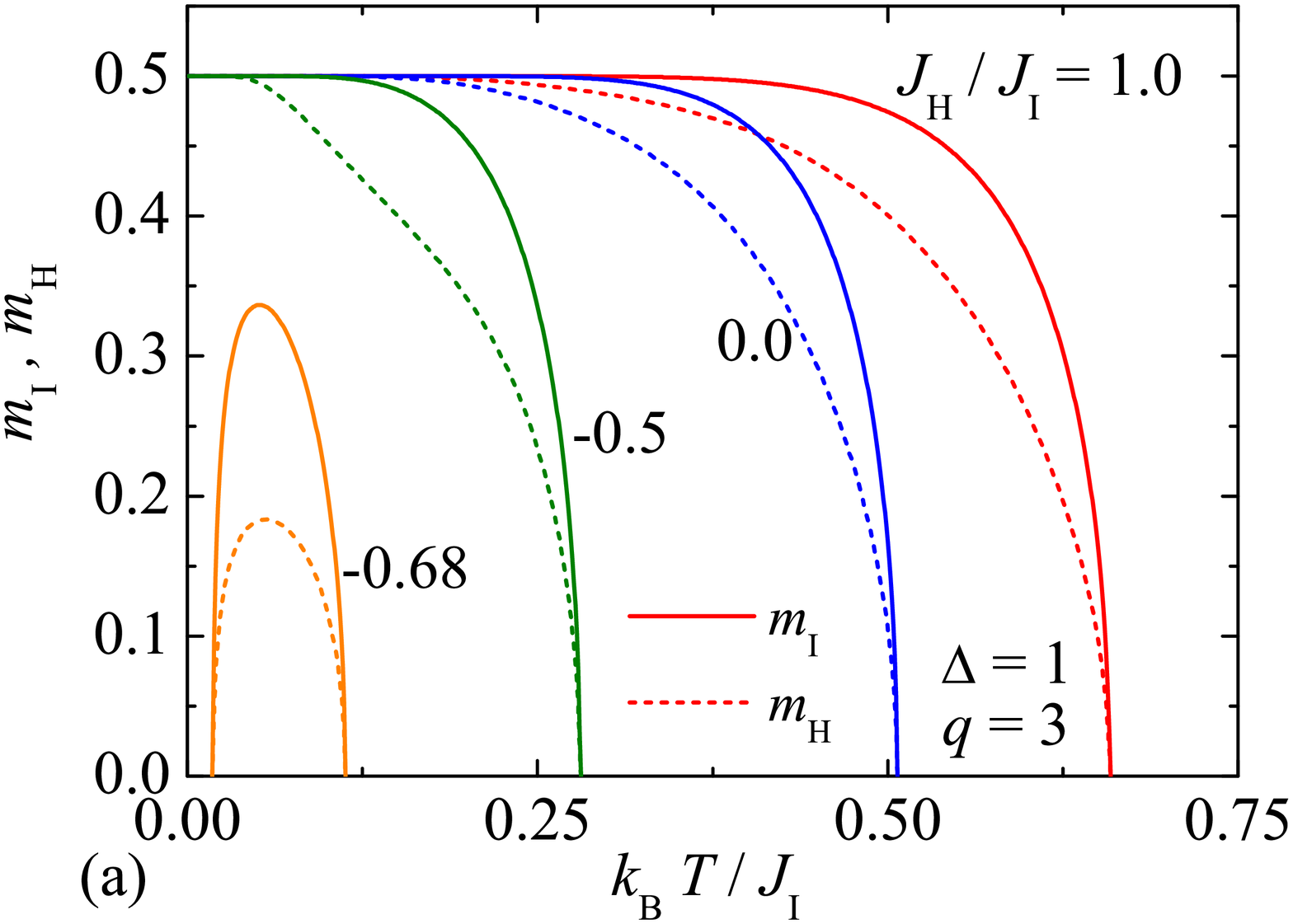}
\includegraphics[width=8cm]{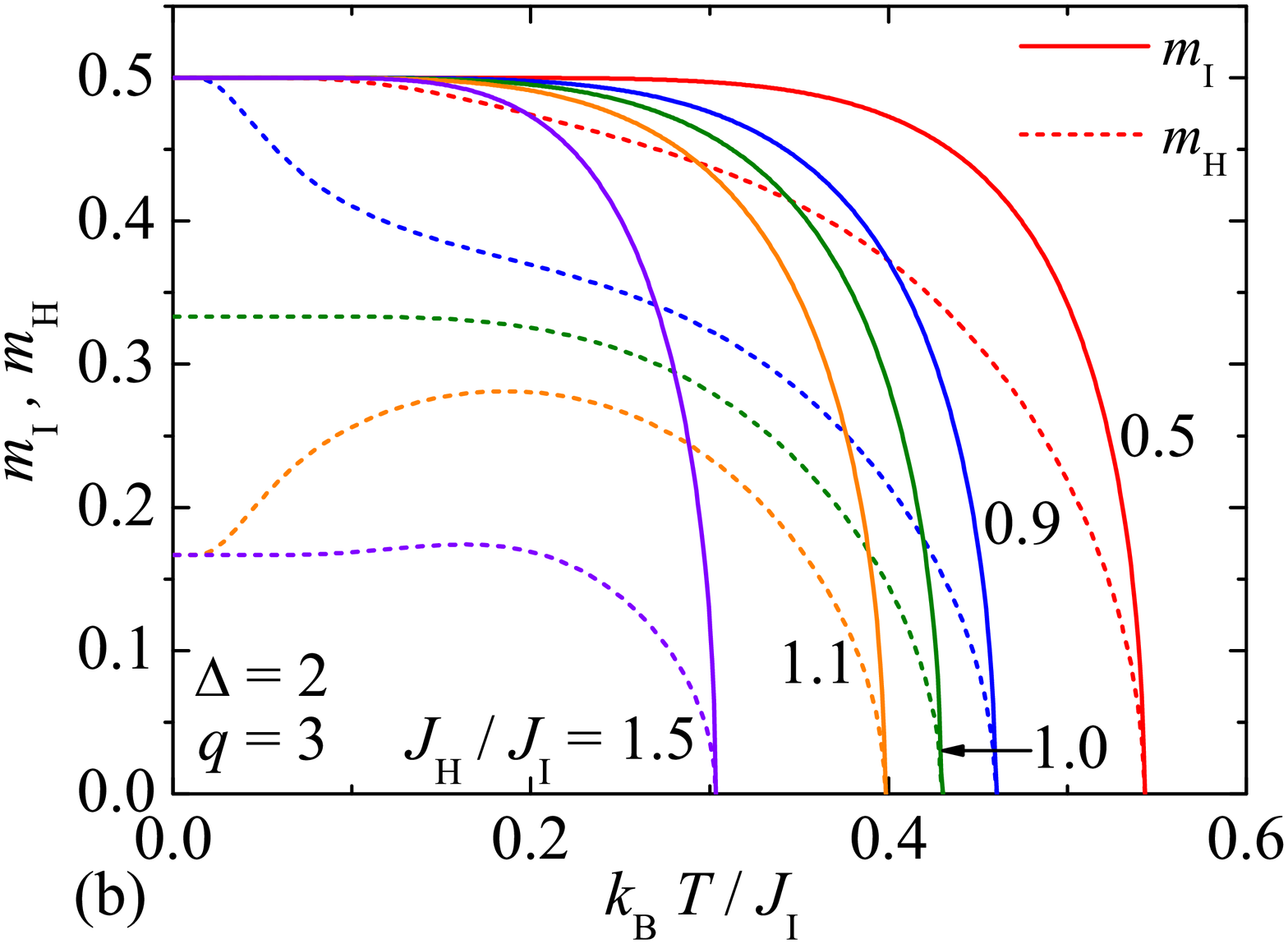}
\includegraphics[width=8cm]{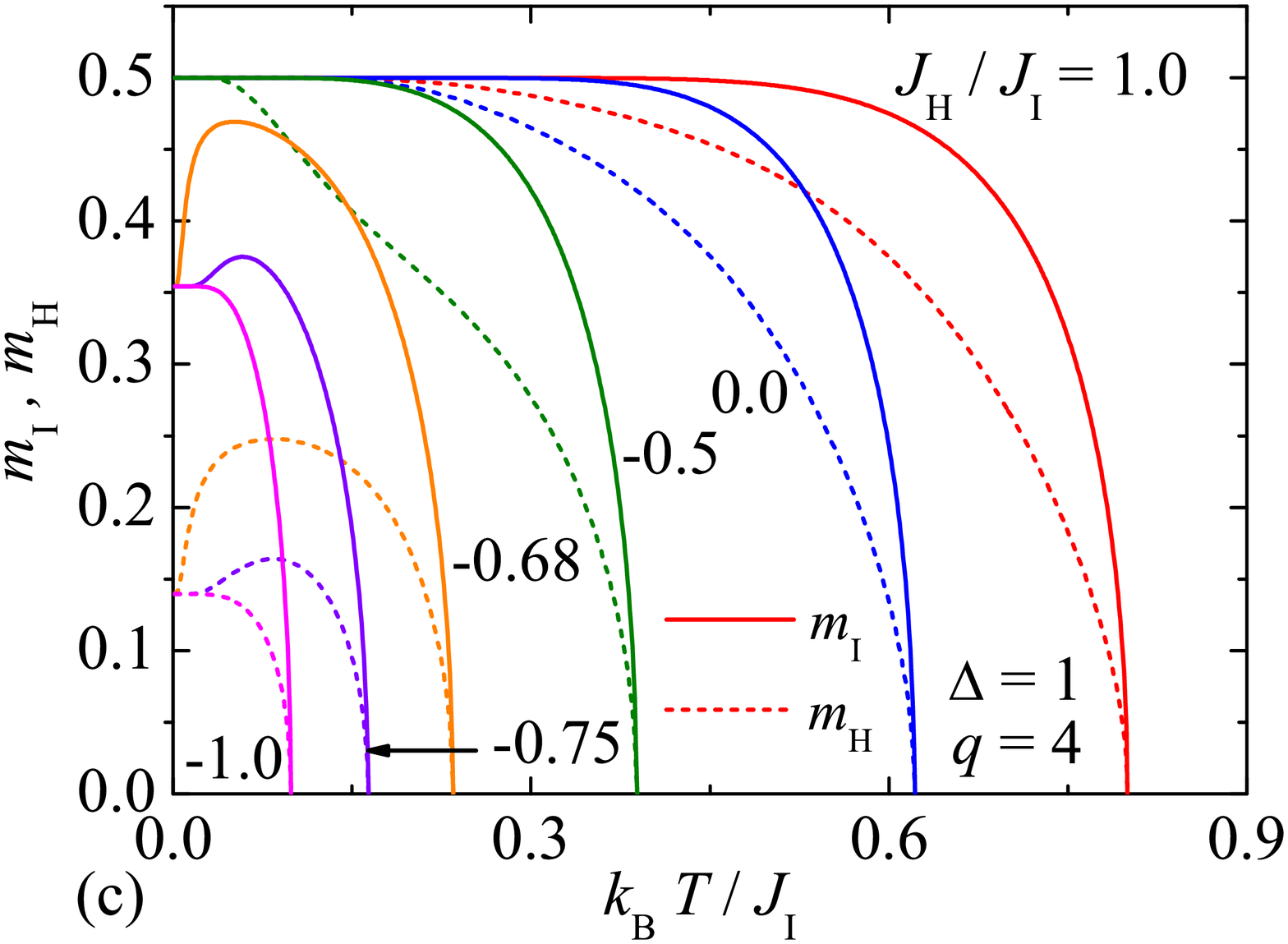}
\vspace*{-0.2cm}
\caption{\small (Color online) Thermal variations of the sublattice magnetizations of the Ising spins (solid lines) and the Heisenberg spins (broken lines) for several values of the interaction ratio $J_{\rm H}/J_{\rm I}$ and a few values of the coordination number and the exchange anisotropy: (a) $q=3$, $\Delta = 1$;
(b) $q=3$, $\Delta = 2$; (c) $q=4$, $\Delta = 1$.}
\label{fig4}
\end{figure}
 
Last but not least, let us discuss typical temperature changes of spontaneous sublattice magnetizations of the Ising and Heisenberg spins, respectively. It can be deduced from Fig. \ref{fig4} that both sublattice magnetizations usually follow the same general trends though the sublattice magnetization of the Heisenberg spins generally falls down more rapidly with an increase in temperature than the sublattice magnetization of the Ising spins. It can be seen from Fig. \ref{fig4}(a) that the critical temperature of the spin-1/2 Ising-Heisenberg model on the triangulated Husimi lattice with the coordination number $q=3$  gradually diminishes as the antiferromagnetic Heisenberg interaction strengthens owing to reinforcement of the geometric spin frustration. The most notable temperature dependence of both sublattice magnetizations can be detected for a sufficiently but not too strong antiferromagnetic Heisenberg interaction ($J_{\rm H}/J_{\rm I} \lesssim -2/3$ for $\Delta=1$), which drives the investigated model to the disordered ground state (\ref{odp}) and simultaneously preserving it in a close vicinity of the phase boundary with the classical ferromagnetic phase (\ref{cfp}). Under these circumstances, the sublattice magnetizations of the spin-1/2 Ising-Heisenberg model on the triangulated Husimi lattice with the coordination number $q=3$ display reentrance with two consecutive critical temperatures in concordance with the finite-temperature phase diagram shown in Fig. \ref{fig3}(b) (see in Fig. \ref{fig4}(a) the dependence for $J_{\rm H}/J_{\rm I} = -0.68$).

To confirm an existence of the quantum ferromagnetic phase (\ref{qfp}) we have depicted in Fig. \ref{fig4}(b) temperature variations of both sublattice magnetizations of the spin-1/2 Ising-Heisenberg model on the triangulated Husimi lattices with the ferromagnetic Heisenberg interaction of the easy-plane type. While the sublattice magnetization of the Ising spins remains fully saturated irrespective of a relative strength of the ferromagnetic Heisenberg interaction, the sublattice magnetization of the Heisenberg spins is fully saturated just if $J_{\rm H}/J_{\rm I}<1$ when considering the special case with the anisotropy parameter $\Delta=2$. Assuming the reverse condition $J_{\rm H}/J_{\rm I}>1$ and $\Delta=2$, the sublattice magnetization of the Heisenberg spins undergoes a quantum reduction to one-third of its saturation value due to a symmetric quantum superposition of three up-up-down spin states of the Heisenberg trimers. This finding evidently corroborates an existence of the quantum ferromagnetic ground state (\ref{qfp}) in overwhelming parameter space with the ferromagnetic Heisenberg coupling of the easy-plane type.

\begin{table}
\vspace{0.0cm}
\caption{Zero-temperature asymptotic limits of the sublattice magnetization of the Ising and Heisenberg spins as a function of the coordination number $q$ of the triangulated Husimi lattices in the highly frustrated region $J_{\rm H}/J_{\rm I}<-2/(2+\Delta)$.}
\vspace{0.0cm}
\begin{center}
\begin{tabular}{|c|c|c|}
\hline $q$ & $m_{\rm I}$ & $m_{\rm H}$  \\ 
\hline $q \leq 3$ & $0$ & $0$ \\ 
\hline $q = 4$ & $0.3543$ & $0.1394$ \\
\hline $q = 5$ & $0.4537$ & $0.1630$ \\
\hline $q = 6$ & $0.4808$ & $0.1659$ \\
\hline \vdots & \vdots & \vdots \\
\hline $q \to \infty$ & $1/2$ & $1/6$ \\ 
\hline 
\end{tabular} 
\end{center}
\label{tab}
\end{table}

Of course, the most challenging is validation of the predicted spectacular quantum order in a highly frustrated region of the spin-1/2 Ising-Heisenberg model on the triangulated Husimi lattices with sufficiently high coordination numbers $q \geq 4$. To illustrate the case, Fig. \ref{fig4}(c) shows typical temperature dependences of both sublattice magnetizations of the spin-1/2 Ising-Heisenberg model on the triangulated Husimi lattice with the coordination number $q=4$ when considering the isotropic Heisenberg interaction $\Delta=1$. It can be readily verified that the sublattice magnetizations of Heisenberg and Ising spins are subject to a quantum reduction caused by local quantum fluctuations. This assertion is consistent with the observed zero-temperature limits of the sublattice magnetizations, which acquire in the highly frustrated region ($J_{\rm H}/J_{\rm I}<-2/3$ for $\Delta=1$) nontrivial asymptotic values. Tab. \ref{tab} summarizes the zero-temperature limits of the sublattice magnetization of Ising and Heisenberg spins for a few selected values of the coordination number $q$, which were obtained by solving a polynomial equation of degree $2q-1$ derived from the exact recursion relation (\ref{eq:14}). It can be understood from Tab. \ref{tab} that the quantum reduction of both sublattice magnetizations gradually shrinks as the coordination number $q$ of the triangulated Husimi lattice increases. The quantum reduction of the spontaneous magnetization of the Ising spins completely disappears in the limit $q \to \infty$ in contrast to the spontaneous magnetization of the Heisenberg spins, which tends towards one-third of its saturation value in the limit $q \to \infty$. Finally, the notable temperature-induced increase in both sublattice magnetizations can be observed in Fig. \ref{fig4}(c) for $J_{\rm H}/J_{\rm I}\lesssim-2/3$, which bears a close relation to a thermal excitation from the ground-state manifold (\ref{odp}) to the classical ferromagnetic phase (\ref{cfp}) accompanied with an increase in the relevant magnetizations (see the curves for $J_{\rm H}/J_{\rm I}=-0.68$ and $-0.75$).

\section{Conclusions}
\label{sec:4}
The present work deals with the exact solution of the spin-1/2 Ising-Heisenberg model on the triangulated Husimi lattices, which has been obtained by the generalized star-triangle mapping transformation and the method of exact recursion relations. In particular, our attention has been focused on a rigorous determination of the ground-state and finite-temperature phase diagrams, which were subsequently corroborated by temperature dependences of the sublattice magnetizations of Ising and Heisenberg spins serving as the relevant order parameters. It has been convincingly evidenced that the spin-1/2 Ising-Heisenberg model defined on two closely related triangulated Husimi lattices may display fundamentally different magnetic behavior even if both lattices may differ from each other just in a connectivity of the same structural triangles-in-triangles unit.

We have furnished rigorous proof that the quantum disorder develops in a highly frustrated region of the spin-1/2 Ising-Heisenberg model on the triangulated Husimi lattices with the coordination number $q=2$ and $3$, while the identical model on the triangulated Husimi lattices with greater coordination numbers $q \geq 4$ exhibits the spectacular quantum order due to the order-by-disorder effect. In addition, it has been demonstrated that the spontaneous magnetizations of the Ising and Heisenberg spins undergo a quantum reduction in the remarkable quantum ordered state, which gradually shrinks with increasing a connectivity of the triangles-in-triangles units. It should be also mentioned that reentrant phase transitions can only be found in the model defined on the triangulated Husimi lattice with the coordination number $q=3$. To conclude, the spin-1/2 Ising-Heisenberg model on the triangulated Husimi lattices represents a novel exactly solved classical-quantum spin model, which has proved its usefulness in elucidating a sought connection between the quantum order-by-disorder effect and the relative strength of local quantum fluctuations varied systematically through the coordination number of the underlying Husimi lattice.

\begin{acknowledgments}
This work was financially supported by the grant of the Slovak Research and Development Agency under the contract \mbox{No. APVV-0097-12} and by the ERDF EU (European Union European regional development fond) grant provided under the contract No. ITMS26220120005 (activity 3.2). 
\end{acknowledgments}

\end{document}